\documentclass[journal]{IEEEtran}
\usepackage{amsmath}
\usepackage{amssymb}
 \usepackage{graphicx}
 \usepackage{subcaption}
 \usepackage{booktabs}
 \usepackage{hyperref}
\usepackage{xcolor}

\ifCLASSINFOpdf
\else
\fi

\begin{document}

\title{Large Speech Model Enabled Semantic Communication}

\author{Yun~Tian,
        Zhijin Qin, ~\IEEEmembership{Senior Member,~IEEE,}
        Guocheng Lv, \\
        Ye Jin,
        Kaibin Huang,~\IEEEmembership{Fellow,~IEEE,}
        and Zhu Han,~\IEEEmembership{Fellow,~IEEE}
\thanks{
This work is supported by the National Key Research and Development Program of China under Grant No. 2023YFB2904300, the National Natural Science Foundation of China under Grant No. 62293484, and Beijing Natural Science Foundation (F251001).
\textit{(Corresponding author: Zhijin Qin.)}}
\thanks{
Yun Tian, Guocheng Lv and Ye Jin are with the School of Electronics, Peking University, Beijing 100091, China. (email: tianyun@stu.pku.edu.cn; lv.guocheng@pku.edu.cn; jinye@pku.edu.cn).}
\thanks{
Zhijin Qin is with the Department of Electronic Engineering, Tsinghua University, Beijing 100084, China, andv with the State Key Laboratory of Space Network and Communications, Beijing, 100084, China. (email: qinzhijin@tsinghua.edu.cn).}
\thanks{Kaibin Huang is with the Department of Electrical and Electronic Engineering, The University of Hong Kong, Hong Kong SAR, China (email: huangkb@hku.hk).}
\thanks{Z. Han is with the Department of Electrical and Computer Engineering
at the University of Houston, Houston, TX 77004 USA, and also with the
Department of Computer Science and Engineering, Kyung Hee University,
Seoul, South Korea, 446-701 (email: hanzhu22@gmail.com).}
}

\markboth{}%
{Shell \MakeLowercase{\textit{et al.}}: Large Model enhanced Video-Speech Semantic Communication}

\maketitle

\begin{abstract}
Existing speech semantic communication systems  mainly based on Joint Source-Channel Coding (JSCC) architectures have demonstrated impressive  performance, but their effectiveness remains limited by model structures specifically designed for particular tasks and datasets. Recent advances indicate that generative large models pre-trained on massive datasets, can achieve outstanding performance arexhibit exceptional performance across diverse downstream tasks with minimal fine-tuning. To exploit the rich semantic knowledge embedded in large models and enable adaptive transmission over lossy channels, we propose a Large Speech Model enabled Semantic Communication (LargeSC) system. 
Simultaneously achieving adaptive compression and robust transmission over lossy channels remains challenging, requiring trade-offs among compression efficiency, speech quality, and latency.
In this work, we employ the Mimi as a speech codec, converting speech into discrete tokens compatible with existing network architectures. We propose an adaptive controller module that enables adaptive transmission and in-band Unequal Error Protection (UEP), dynamically adjusting to both speech content and packet loss probability under bandwidth constraints. Additionally, we employ Low-Rank Adaptation (LoRA) to finetune the Moshi foundation model for generative recovery of lost speech tokens. Simulation results show that the proposed system supports bandwidths ranging from 550 bps to 2.06 kbps, outperforms conventional baselines in speech quality under high packet loss rates and achieves an end-to-end latency of approximately 460 ms, thereby demonstrating its potential for real-time deployment.
\end{abstract}


\IEEEpeerreviewmaketitle

\section{Introduction}
Driven by recent advances in Artificial Intelligence (AI) and the increasing demand for intelligent next-generation communication systems, semantic communication has attracted significant attention. Semantic communication aims to ensure accurate semantic delivery rather than pursuing bit-level precise communication \cite{qin2021semantic,AIempowered}. 
Semantic communication systems exploit the feature extraction and generative capabilities of AI models to convey the underlying semantics of the source, thereby enhancing communication efficiency through the integration of computational resources \cite{ComputingNetworks}.
However, existing semantic communication systems are often constrained by their reliance on specific datasets or tasks, requiring costly model redesign or retraining for new application. In contrast to task-specific semantic communication systems, semantic communication systems employing large pre-trained models exhibit superior performance across diverse tasks through minimal fine-tuning. 
Motivated by the remarkable performance of large language models (LLMs) (e.g., ChatGPT \cite{achiam2023gpt} and DeepSeek \cite{guo2025deepseek}) and image generative models (e.g., Stable Diffusion \cite{rombach2022high}), recent studies have explored the use of large models to enhance semantic communication, particularly in text and image transmission \cite{wang2024uses,jiang2024large,nam2024language}.
Despite this progress, the application of large models in adaptive low-bitrate speech transmission remains underexplored, and yet holds significant potential for enabling robust and efficient speech communication over bandwidth-constrained and packet loss channels.
\subsection{Piror Work}
The primary objective of existing speech semantic communication systems is to extract speech-relevant features for bandwidth compression while ensuring robust transmission over various channels.
Current end-to-end speech semantic communication systems can be categorized based on the type of transmitted semantic features:
text-related and acoustics-related. Transmitting text-related semantic features enables ultra-low bitrate communication while preserving the semantic fidelity of speech through generative reconstruction at the receiver. 
DeepSC-ST \cite{weng2023speech} transmits speech recognition-related features to enable speech synthesis at the receiver.  However, transmitting only linguistic features is insufficient for high-fidelity reconstruction.  In \cite{efficientspeech}, additional speech related information for duration, pitch, and power are directly transmitted to guide the speech synthesis.
In SyncSC \cite{syncSC}, visual information is leveraged to guide the generation of prosodic attributes such as duration and pauses, thereby enhancing the realism and expressiveness of the synthesized speech.
Although the transmission of text-related features enables significant bandwidth compression, the synthesized speech often suffers from low quality and synchronization challenges. These approaches struggle to retain rich prosodic and paralinguistic information, such as pauses and emotions. Moreover, the two-stage process of speech-to-text and text-to-speech conversion typically incurs delays on the order of seconds, which severely limits their applicability in real-time communication scenarios. 

An alternative approach transmits acoustic features through autoencoder-structured Deep Neural Networks (DNNs) for speech compression and reconstruction. 
In DeepSC-S \cite{weng2021semantic}, a semantic-channel coder with an attention mechanism is jointly optimized for time-domain speech signals. A low-complexity, fully convolutional semantic coder is proposed in \cite{lowSC}, while DeepSC-TS \cite{swinSC} transmits multi-level semantic features to improve speech quality.
However, the high dimensionality of multi-channel floating-point features generated by convolutional networks hinders bitrate reduction, and the large coding window introduces additional latency. For digital semantic transmission, Soundspring \cite{soundspring} addresses these issues by compressing audio into quantized discrete tokens, enabling lower bitrates and compatibility with existing communication systems. Inspired by this, our proposed system adopts discrete speech tokens to represent speech semantics, enabling adaptive compression and in-band unequal error protection based on content and channel conditions. By integrating a large pre-trained speech model, the system enhances semantic representation and generative reconstruction, is designed with the goal of supporting real-time communication scenarios under lossy and bandwidth-constrained conditions.

\begin{figure*}
    \centering
\includegraphics[width=0.8\linewidth]{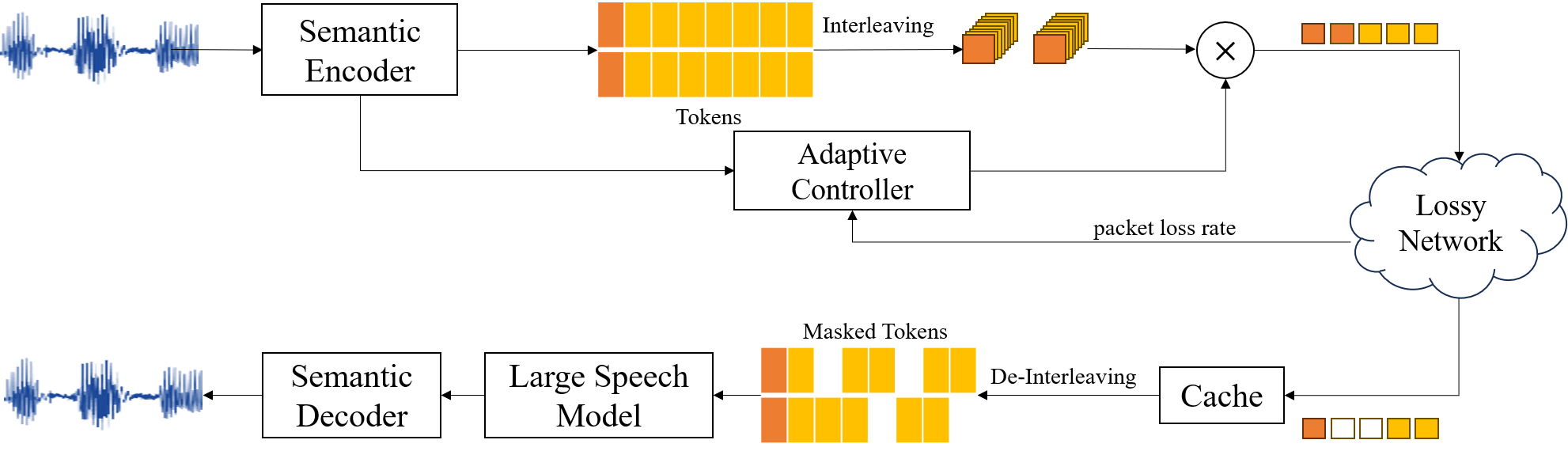}
    \caption{The proposed framework of large speech model enabled semantic communication system.}
    \label{figure_1_framework}
\end{figure*}

Improving the robustness of speech transmission over unreliable channels is essential for semantic communication. 
Channel coding or error control techniques improve the transmission robustness of speech in various  channels. 
In real-time scenarios, Forward Error Correction (FEC) is generally preferred over retransmission due to its lower Round-Trip Time (RTT). Traditional FEC schemes based on linear block codes, such as Reed-Solomon (RS) codes, can achieve reliable performance under low error rates. However, their effectiveness significantly degrades under low signal-to-noise ratio (SNR) or high packet loss conditions. In addition, these methods require a careful balance between redundancy rate and available bandwidth under varying channel conditions. Recent advances in JSCC offer improved resilience to channel noise, especially under low-SNR conditions. By jointly optimizing semantic-channel coding, these methods can better maintain end-to-end performance in challenging environments.
Over \cite{weng2023speech,weng2021semantic}, linear layers are used as a channel coder to transform the dimensionality of semantic features, enabling joint training under varying physical-layer SNR conditions.  However, integrating analog semantic signals with existing digital communication systems remains challenging. To address this issue, digital semantic communication is proposed based on modulation and quantization of semantic symbols. 
On the modulation side, \cite{tung2022deepjsccq,bo2024joint} incorporate digital modulation into semantic communication by mapping semantic symbols onto constellation points. \cite{zhang2024analog} proposes a trainable multi-order and variable-rate modulation scheme compatible with conventional digital transmission. In addition, \cite{park2024joint} develops a channel-adaptive JSCC framework that employs ternary robust demodulation and adaptive modulation to preserve task performance with reduced latency. For the quantization of semantic features, VQ-DeepSC \cite{fu2023vector} employs multi-scale vector quantization to map semantic features into discrete codebook indices for bit-level transmission. \cite{guo2024digital} proposes a learned nonlinear quantization mechanism that narrows the performance gap between digital semantic communication and analog JSCC.
\cite{bao2025sdac} presents a semantic digital–analog converter enabling bidirectional transformation between semantic features and digital bits. \cite{chen2025low} further develops a residual vector quantization GAN framework for low-bitrate digital semantic communication.
However, most existing studies focus on robustness at the physical layer under varying SNR conditions, while packet-level optimization at higher layers has been relatively underexplored.

To improve robustness under packet loss channel, \cite{syncSC} proposes a packet-level coding method for video semantics.  Another line of work focuses on packet loss concealment, where loss-resilient semantic representations are transmitted and the receiver performs predictive recovery. In this context, the packet loss problem is modeled as a masked language modeling task. A bidirectional Transformer encoder is employed to recover lost words in \cite{syncSC}, and to predict lost audio tokens in \cite{soundspring}. However, these approaches face challenges in real-time communication scenarios. The bidirectional transformer relies on both past and future context, which introduces latency and makes it less suitable for streaming applications. To address this limitation, we propose an packet loss concealment method based on a large pre-trained speech model with an autoregressive structure. By depending only on past received packets, it is inherently better suited for real-time applications.

Large AI models have demonstrated strong potential in semantic communication systems. Most existing large AI models are generative models with autoregressive structures, trained on massive datasets. These models have attracted increasing attention due to their strong contextual understanding, generation capabilities, and state-of-the-art performance across a wide range of downstream tasks \cite{chen2024big}. In semantic communication, large model enhanced systems typically leverage these capabilities to improve transmission performance and reduce bandwidth overhead. For instance, a privacy-preserving semantic communication framework using multimodal LLMs is proposed in \cite{cao2024multimodal}, where text serves as a unified representation to protect sensitive information through few-shot learning. An end-to-end text transmission system is proposed in \cite{wang2024uses}, incorporating an LLM-based semantic decoder. For image transmission, SAMSC \cite{jiang2024large} applies the Segment Anything Model (SAM) to partition images into semantic regions, reducing communication overhead. LSC \cite{nam2024language} proposes a language-guided semantic framework that enhances robustness over noisy channels by leveraging heterogeneous Text2Image and Image2Text knowledge. However, existing studies primarily focus on language and vision modalities, with limited exploration of speech as a source. The integration of large models for speech semantic communication remains an open and promising direction.

\subsection{Motivation and Contributions}
Despite the rapid development of large language and vision models in semantic communication, existing speech semantic communication systems rarely exploit the generative capabilities of large pre-trained models.
To address this gap, we propose a speech semantic communication framework enhanced by a large model, LargeSC,  which represents speech using discrete tokens to enable low-bitrate transmission and formulates packet loss concealment as a masked token prediction task. The system integrates adaptive compression and robust transmission strategies to balance compression efficiency, reconstruction quality, and resilience to packet loss in real-time communication and lossy channel. 
At the transmitter, an adaptive compression and in-band unequal error protection strategy is applied based on both channel status and target bandwidth.  At the receiver, built upon an autoregressive large speech model, LargeSC performs packet loss concealment. The main contributions are as follows:

\begin{enumerate}
\item \textbf{Semantic-aware adaptive compression:} We develop an extended version of the pre-trained Mimi model to serve as the semantic coder for 24 kHz speech, supporting variable bitrates from 550 bps to 2.06 kbps at a 12.5 ms frame rate.

\item \textbf{Loss-aware in-band unequal error protection:} We design an adaptive controller module that dynamically adjusts semantic compression based on speech content importance and channel conditions. The module allocates bandwidth efficiently while applying enhanced protection to critical tokens under lossy networks.

\item \textbf{Real-time transmission capability:} Leveraging the autoregressive structure of the large model, our system performs causal packet loss concealment at the receiver. This results in an end-to-end latency of approximately 460 ms, demonstrating potential for real-time communication scenarios.

\end{enumerate}

\subsection{Organization}
The rest of this paper is organized as follows. The framework of large speech model enabled semantic communication system is introduced in Section \ref{sec:II}. Section \ref{sec:III} describes the details of the proposed model. The simulation results are presented in Section \ref{sec:IV}. Section \ref{sec:V} discusses potential applications, limitations, and future challenges. Finally, Section \ref{sec:VI} concludes the paper.

\section{System Model}
\label{sec:II}

In this section, we introduce the system model of the proposed LargeSC, as illuistrated in Fig. \ref{figure_1_framework}. The speech coder, built upon the Mimi \cite{moshi} model, performs the conversion between speech signals and discrete tokens. An adaptive controller module is designed to perform semantic-aware compression and in-band unequal error protection, dynamically adjusting transmission according to token importance and channel conditions. The large speech model, Moshi \cite{moshi} is employed at the receiver to predict lost packets through generative inference.

\subsection{Semantic Encoder}
In the proposed system, the speech transmission task is reformulated as a token-level transmission problem. The input of the semantic encoder is speech chunks in time domain, denoted as $\boldsymbol{S} = [s_1, s_2, ...,s_n,..,s_{N}]$, where $N$ is the total number of speech frames. For real-time speech transmission, each speech chunk $s_n$ is encoded into $N_Q$ tokens by the semantic encoder, which can be represented as
\begin{equation}
    x_n= \text{Encoder}(s_n),
\end{equation}
where the token $x_n = [x_{n_1},x_{n_2},...,x_{n_{N_Q}}]$ refers to the index set of quantized speech feature representations in the codebook.
We adopt Mimi as the speech codec, configured with $N_Q$ codebooks. 
The encoder of Mimi consists of convolutional layers and transformer blocks that extract a latent feature vector $z_n$ from the input speech chunk $s_n$. This latent feature is then quantized using Residual Vector Quantization (RVQ) to produce the quantized representation $z_{q_n}$. 
RVQ quantizes the latent vector $z_{q_n}$ in a multi-stage recursive manner. At stage  $i$, the residual is defined as:
\begin{equation}
r_{i,n} =
\begin{cases}
z_n, & i = 1, \\[4pt]
r_{i-1,n} - Q_{i-1}(r_{i-1,n}), & i \ge 2.
\end{cases}
\end{equation}
where $Q_i(\cdot)$ denotes the $i$-th quantizer. Each residual represents the reconstruction error after subtracting the contributions of all previous quantizers. The final quantized representation is computed by accumulating the outputs of all RVQ stages:
\begin{equation}
z_{q_n} = \sum_{i=1}^{N_q} Q_i(r_{i,n}).
\end{equation}
After training the codebooks, the quantized indices from the $N_Q$ quantizers are used as discrete tokens $x_n$ for transmission. Specifically, $x_{n_0}$ is referred to as the \emph{semantic token}, which carries high-level semantic content, while $x_{n_1}$ through $x_{n_7}$ are termed \emph{acoustic tokens}, containing finer acoustic details.

\subsection{Adaptive Controller}
After encoded by the speech encoder, each speech chunk is represented by a fixed number of tokens. However, fixed-length token representations may lead to inefficient bandwidth usage, and the loss of certain tokens can significantly degrade transmission quality. To address these challenges, we design an adaptive controller that considers the following factors:

\begin{itemize}

\item \textbf{Causal token importance:} Due to the residual quantization structure, tokens from lower-ranking quantizers carry more critical semantic information. Moreover, the autoregressive decoding process imposes a strict causal dependency, making earlier tokens critical for accurately predicting subsequent ones.

\item \textbf{Content-dependent token demand:} Different speech segments require varying numbers of tokens; for instance, silence or low-energy regions can be effectively represented with fewer tokens.

\item \textbf{Importance-aware error protection:} Under fluctuating channel conditions, it is essential to prioritize the protection of important tokens and introduce redundancy selectively.
\end{itemize}
\begin{figure*}[!ht]
    \centering
    \includegraphics[width=0.7\linewidth]{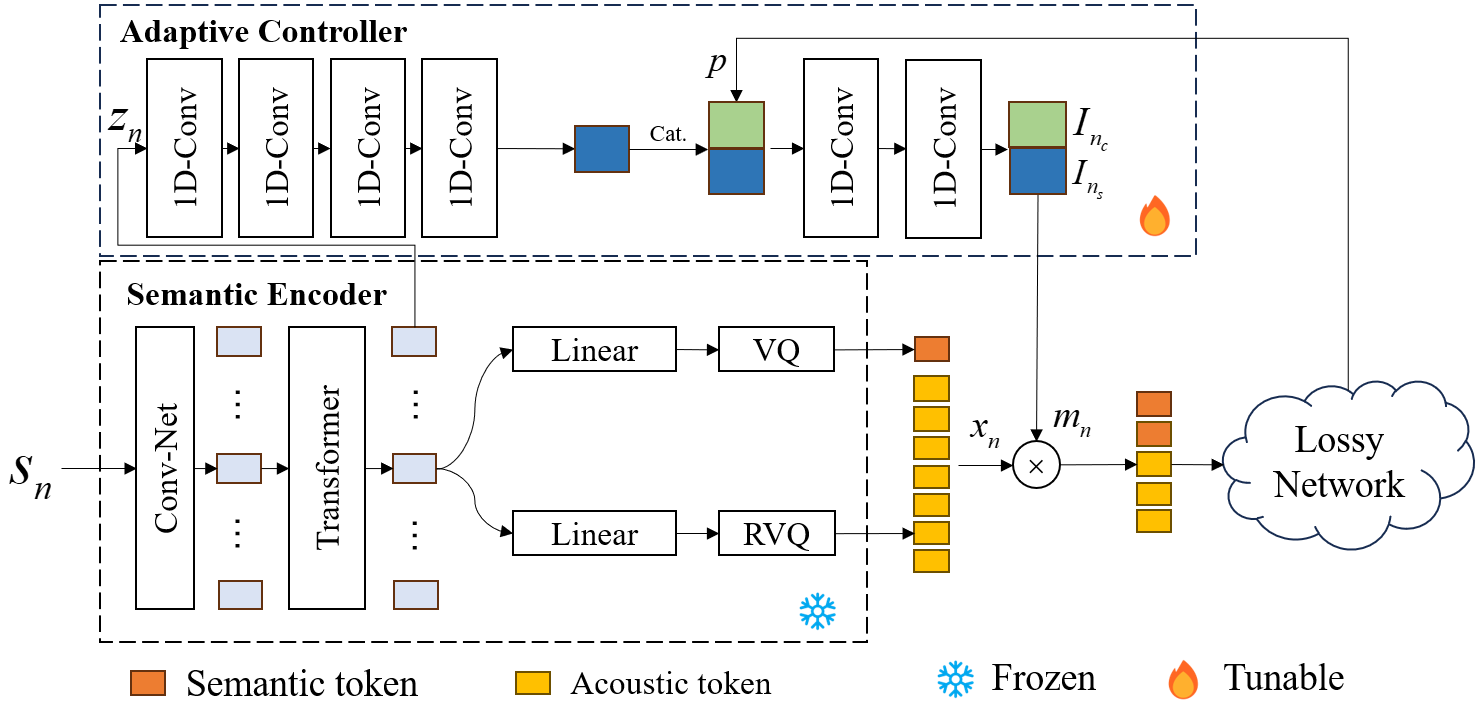}
    \caption{Architecture of the semantic encoder and the adaptive controller.}
    \label{fig_2}
\end{figure*}
Therefore, the output of the speech encoder is subject to adaptive compression and in-band unequal error protection, guided by both the semantic importance of tokens and channel conditions.

The unquantified feature $z_n$ of a speech chunk and  the corresponding packet loss rate $p$ are jointly used to compute a semantic importance map $I_n$, which can be expressed as
\begin{equation}
    I_n = \text{Controller}_{\alpha}(z_n,p),
\end{equation}
where $\alpha$ denotes the trainable parameters of the adaptive controller, and $I_n = [I_{n_s}, I_{n_c}]^T\in[0,1]$. 
The component $I_{n_s}$ reflects the semantic importance of the speech content—segments with richer content are assigned higher values, corresponding to lower compression rates. 
The component $I_{n_c}$ captures the impact of the lossy channel. More adverse conditions result in higher values, indicating the need for increased redundancy. 
The adaptive controller is trained to achieve a balanced trade-off between compression and redundancy.
The importance vector $I_n$ is then mapped to a token-level mask $m_n \in \{0,1,2\}^{N_Q}$ using a step function $H^j(i)$ in \cite{VBRVQ}, defined as:
\begin{equation}
    \begin{aligned}
    m_{n_s} &= [H^1(L\cdot I_{n_s}),...,H^{N_Q}(L\cdot I_{n_s})] \\
    m_{n_c} &= [H^1(L\cdot I_{n_c}),...,H^{N_Q}(L\cdot I_{n_c})] \\
    m_{n} &= m_{n_s} +m_{n_c},
    \end{aligned}
\end{equation}
where $m_{n_s}$ and $m_{n_c}$ are binary masks derived from $I_{n_s}$ and $ I_{n_c}$, respectively. $L$ denotes the number of quantizers used, which is proportional to the target bitrate. The step function $H^j(i)$ is defined as follows:
\begin{equation}
    H^j(i) = \begin{cases}
        1 & \text{if } j \leq i, \\
        0  & \text{if } j > i.
    \end{cases}
\end{equation}
The resulting mask $m_n$ determines whether each token in $x_n$ should be discarded ($m_{n_i}=0$), transmitted once ($m_{n_i}=1$), or repeated for redundancy ($m_{n_i}=2$), yielding the final transmitted sequence:

\begin{equation}
x_{n_{\text{trans}}} = m_n \cdot x_n.
\end{equation}
To further mitigate the impact of burst losses and prevent the complete loss of tokens for a given chunk, interleaving is applied across adjacent speech chunks. Specifically, the tokens at even indices are swapped between adjacent chunks before transmission.

\subsection{Large Speech Model and Semantic Decoder}
After transmission over a lossy channel and de-interleaving, packet losses are reflected as masked tokens in the received sequence.  The masked tokens, denoted as $\tilde{x}$, are recovered by an autoregressive large speech model. It can be expressed as
\begin{equation}
    \hat{x}_n= \text{LM}_{\beta}(\tilde{x}_n, \hat{x}{<n}),
\end{equation}
where $\beta$ denotes the trainable parameters of the Large Model (LM). The model takes the partially received tokens $\tilde{x}_n$ at the current step and the previously reconstructed tokens $\hat{x}{<n}$ as inputs to autoregressively estimate the missing tokens in $x_n$.
In our system, we employ a fine-tuned Moshi \cite{moshi} speech-to-speech model to perform token prediction. 
Finally, the estimated $\hat{x}_n$ is decoded to the reconstructed speech chunk $\hat{s}_n$ through the semantic decoder, which can be represented as
\begin{equation}
    \hat{s}_n = \text{Decoder}(\hat{x}_n).
\end{equation}
Therefore, the semantic decoder integrates generative reconstruction with causal inference, enabling robust recovery of speech content under lossy transmission conditions.

\section{Proposed LargeSC Architecture}
\label{sec:III}
This section presents the overall architecture of the proposed LargeSC. The system is composed of four key modules: a speech encoder that converts speech into discrete tokens, an adaptive controller for adaptive compression and protection, a large model for token loss concealment, and a speech decoder that converts the predicted tokens back into waveform.
\subsection{Semantic Coder}
Mimi is employed for encoding and decoding of speech chunks. It consists of an encoder, a set of quantizers, and a decoder. In traditional semantic communication systems such as \cite{weng2021semantic}, speech semantics are represented as continuous features, which are suitable for analog communication. Mimi approximates these continuous features using vectors from learnable codebooks, and represents speech tokens as the index set of the nearest codebook vectors. In contrast, the continuous semantic features of speech are approximated by vectors in learnable codebooks, and speech tokens are the index set of the nearest vector. Using discrete tokens for speech offers several advantages. It improves compatibility with existing layered digital communication architectures. Moreover, since the input format of generative large models is typically token-based, this representation not only compresses the speech signal for efficient transmission, but also facilitates downstream post-processing tasks.

The encoder of Mimi functions as a tokenizer that represents speech using a learned codebook of discrete tokens. 
A speech chunk $s_n$ is first projected into a latent representation through four residual convolutional  blocks with weight normalization. For 24kHz speech, these convolutions downsample the signal to a latent feature with a temporal resolution of 12.5 frames per second. This feature is then processed by an 8-layer Transformer to enhance the quality and expressiveness of the representation. The resulting latent feature $z_n$ is quantized by Residual Vector Quantization (RVQ). Specifically, $N_Q=8$ quantizers are used, with the first vector quantizer producing a semantic token and the remaining seven quantizers generating acoustic tokens. The number of generated tokens is significantly smaller than that of the original waveform samples. For 24kHz input, the speech encoder generates outputs tokens at a rate of 12.5Hz.  
Each quantizer has a codebook size of 2048, requiring $\log_22048=11$ bits per token. Therefore, the output of the speech encoder is a  compressed bitstream at 1.1kbps. The decoder of Mimi reconstructs the speech waveform from the discrete token sequence.  Its architecture mirrors that of the encoder but replaces convolutional layers with transposed convolutions to perform upsampling.

\begin{figure*}[!t]
    \centering
    \includegraphics[width=0.65\linewidth]{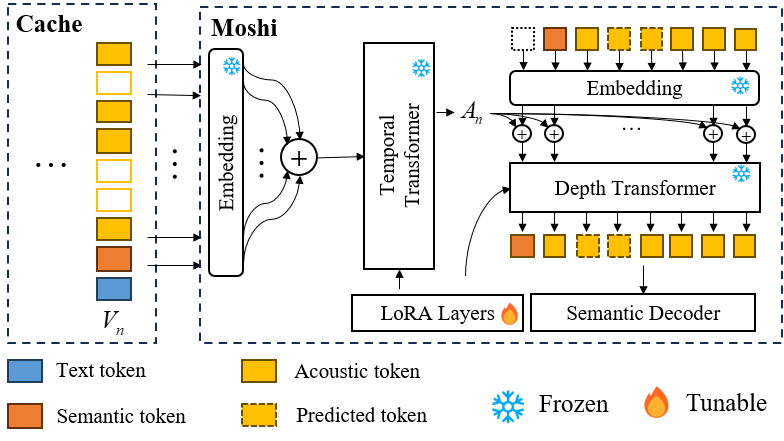}
    \caption{Architecture of Moshi large speech model as packet loss concealment.}
    \label{fig_3}
\end{figure*}

\subsection{Adaptive Controller with Unequal Error Protection}

The speech encoder outputs a fixed-bitrate token sequence, but transmitting all tokens is inefficient.
Speech content is often temporally sparse and semantically unbalanced—segments such as silence or low-information regions require fewer tokens, while semantically rich regions deserve finer representation.
Furthermore, due to the causal nature of autoregressive generation at the receiver, earlier tokens have a greater impact on reconstruction quality and therefore require stronger protection. To address these issues, we design an adaptive controller with UEP, which performs adaptively semantic-aware compression and channel-aware redundancy allocation. 

\subsubsection{Algorithm} The adaptive controller module learns to determine the number of codebooks to transmit and which tokens to repeated, conditioned on the speech content and packet loss rate. This dynamic selection enables flexible bitrate adjustment.
The token selection and protection strategy can be formulated as the following optimization problem:
\begin{equation}
\begin{aligned}
& \underset{\alpha,m_n}{\text{min}} 
& & \sum_{i=1}^{N_Q} m_{n_i} \\
& \text{s.t.}
& & \sum_{i=1}^{N_Q} m_{n_i} \leq 2N_Q, \\
&&& m_{n_i} \in \{0, 1, 2\}, \quad \forall i,
\end{aligned}
\label{loss_bitrate}
\end{equation} 
where $m_{n_i}$ denotes the masking level of the $i$-th token in chunk $n$. The optimization objective of is to minimize the bandwidth usage and add the redundancy of more important tokens.

\subsubsection{Structure} As shown in Fig. \ref{fig_2}, the adaptive controller consists of two convolutional sub-networks, namely the semantic importance extraction module and the channel adaptation module. The semantic importance extractor consists of four one-dimensional convolutional layers with a kernel size of 3 and Snake activation \cite{snake}. These layers progressively reduce the dimensionality of the 512-dimensional latent feature $z_n$ to 256, 128, 64, and finally 1, producing the semantic importance score $I_{{sem}_n}$. The output $I_{{sem}_n}$ is concatenated with the packet loss rate $p$ and fed into the channel adaptation module, which consists of two one-dimensional convolutional layers with a kernel size of 1. After fusing the semantic-aware and channel-aware features, a sigmoid activation function is applied to generate the final importance vector $I_n=[I_{n_s},I_{n_c}]^T$. The importance vector $I_n$ is then mapped to the token-level mask $m_n$ using a step function $H^j(i)$. However, since the step function is non-differentiable, the continuous approximation $H_{soft}^j(i)$ as proposed in \cite{VBRVQ} is adopt to enable gradient-based optimization during training. 

\begin{equation}
    H_{soft}^j(i) = \frac{1}{2\tau} \log \left( \frac{\cosh(\tau(i-j))}{\cosh(\tau(-i+j+1))} \right) +\frac{1}{2},
\end{equation}
where $\tau$ is a hyperparameter controlling the sharpness of the approximation. 
During training, we adopt the straight-through estimator (STE) \cite{liu2022nonuniform} to enable gradient backpropagation through the discrete mask. Specifically, the soft mask \( m^{\text{soft}}_{n_s} \) is computed as:
\begin{equation}
m^{\text{soft}}_{n_s} = \Bigl[ H_{\text{soft}}^1(L \cdot I_{n_s}), \ldots, H_{\text{soft}}^{N_Q}(L \cdot I_{n_s}) \Bigr],
\end{equation}
and the final mask $\tilde{m}_{n_s}$ used for forward computation is given by:
\begin{equation}
\tilde{m}_{n_s} = m^{\text{soft}}_{n_s} + \text{sg}\Bigl( m_{n_s} - m^{\text{soft}}_{n_s} \Bigr),
\end{equation}
where \( \text{sg}(\cdot) \) denotes the stop-gradient operator, which blocks gradient flow during backpropagation.

\subsection{Finetuned Large Speech Model for Token Prediction}
Moshi is a fundation model trained on large-scale speech-text datasets, and designed to support full-duplex human-machine spoken dialogue. 
In our system, Moshi is fine-tuned to predict lost speech tokens along both the temporal axis and across the residual vector quantization (RVQ) depth, using only the previously received tokens.
Unlike the bidirectional masked language model employed in \cite{soundspring}, which relies on non-autoregressive architectures. 
We adopt an autoregressive speech-to-speech structure. This causal property makes it inherently more suitable for streaming and real-time communication, as it avoids reliance on future context during inference and enables efficient token-by-token generation for loss concealment.

Moshi adopts the architecture of RQ-Transformer \cite{lee2022autoregressive}, which comprises two cascaded modules: a temporal transformer and a depth transformer. At each time step $n \leq N$, the input sequence is denoted by $V_n=[V_{n,1},...,V_{n,K}]$, which represents a subset of speech tokens spanning the current and previous time steps. For $1\leq n \leq N $, the previous $(V_0,...,V_{n-1})$ is first encoded by the temporal transformer to obtain the context representation, which can be expressed as
\begin{equation}
    A_n = \text{En}_\text{temp}(V_0,...,V_{n-1}),
\end{equation}
where $A_n\in R^{1024}$ is the temporal context vector, and for $1\leq k\leq K$, the initial values $V_{0,k}=0$. For each time step $n$, the tokens quantized by RVQ at depth levels $1<k\leq K$ are autoregressively decoded by the depth transformer. The previous $(V_{n,1},...,V_{k-1})$, along with the temporal context $A_n$, are used to predict the logits at depth $k$, which can be expressed as
\begin{equation}
    l_{n,k}=\text{De}_\text{depth}(A_n,V_{n,1},...,V_{n,k-1}),
\end{equation}
where $l_{n,k}$ is the predicted logits corresponding to the token $V_{n,k}$. In particular, $l_{n,1}=\text{Linear}(A_n)$ is used to predict the aligned  text token $W_n$ via a linear transformation of the temporal context.
The probability distribution of $V_{n,k}$ is 
obtained by applying softmax to the logits $l_{n,k}$.
As described in \cite{moshi}, a one-step temporal delay is introduced between the semantic token and the acoustic tokens during training and inference, which has been shown to improve reconstruction quality. 
Based on this design, for all time steps $n$, the input subsequences $V_{n,k}$ to Moshi is expressed as follow:
\begin{equation}
    \begin{cases}
    V_{n,1}=W_{n}, &  \\
    V_{n,2}=x_{n_1}, & \\
    V_{n,1+k}=x_{{n-1}_k},       & n\geq 2, 1<k\leq N_{Q},    \\
    V_{n,k} =0,    & n<2,1<k\leq N_Q,
\end{cases}
\end{equation}
where $W_n$ denotes the aligned text token and only participate in training stage. $x_{n_k}$ is the $k$-th quantized token at time step $n$.
As shown in Fig. \ref{fig_3}, the lost tokens in $V_n$ are reconstructed autoregressively, conditioned on all previous tokens $V_i$ with $i<n$. The generation probability for $V_{n,k}$ is given by:
\begin{equation}
\begin{aligned}
    P(V_{n,k}\mid V_{i<n})&=P(V_{n,k}|V_0,...,V_{n-1},V_{n,1},...,V_{n,k-1})\\
    &=\text{softmax}(l_{n,k}).
\end{aligned}
\end{equation}

However, Moshi \cite{moshi} is designed for human-machine spoken dialogue and does not directly support the functionality required for predicting lost tokens in our semantic communication system. 
To address this limitation while maintaining computational efficiency, we adopt Low-Rank Adaptation (LoRA) \cite{hu2022lora} to fine-tune Moshi. 
\begin{figure*}[!htb]
    \centering
    \begin{subfigure}{0.45\textwidth}
        \centering
        \includegraphics[width=\textwidth]{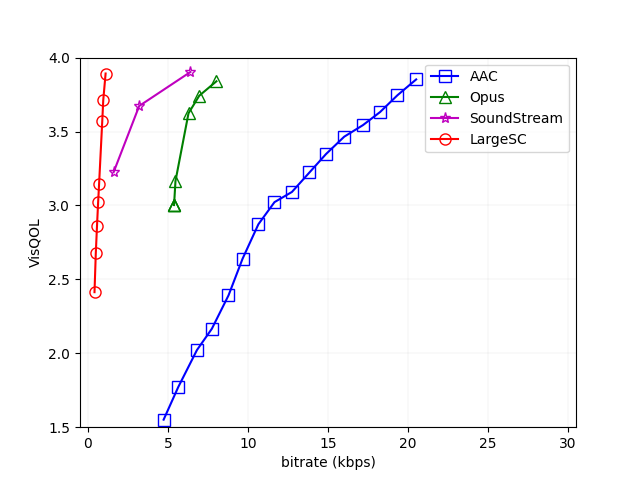}
        \caption{}
    \end{subfigure}
    \begin{subfigure}{0.445\textwidth}
        \centering
        \includegraphics[width=\textwidth]{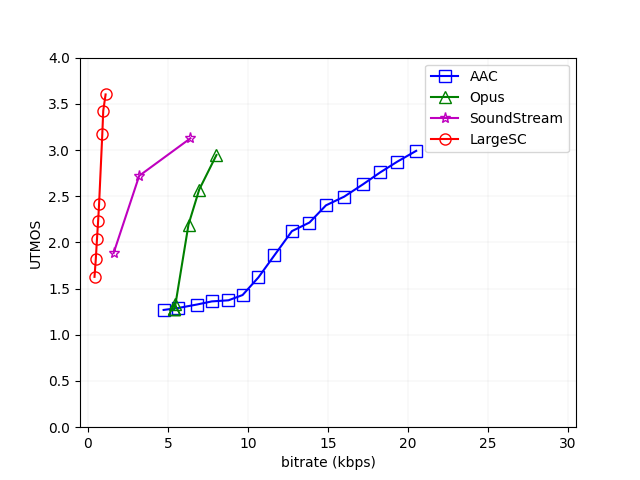}
        \caption{}
    \end{subfigure}
       \caption{Audio quality performance of the proposed system and competing methods with different bitrates (kbps): (a) VisQOL, (b) UTMOS.}
       \label{fig_4}
\end{figure*}
Specifically, LoRA inserts trainable low-rank matrices into the Transformer’s linear layers without modifying the original weights. The adapted weight matrix is expressed as:
\begin{equation}
    W_0 + \Delta W = W_0 + BA.
\end{equation}
Here $W_0\in R^{d\times h}$ represents the frozen pretrained weights. $\Delta W = BA$ is the trainable adaptation, 
where $B\in R^{d\times r}, A \in R^{r\times h}$ 
with rank $r\ll \text{min}(d,h)$. During training, only the parameters of $A$ and $B$ in LoRA layers are updated, while the parameters $W_0$ of the original model remain fixed.

\subsection{Loss Function}
In our system, the speech transmission task is modeled as the transmission of discrete speech tokens. The overall optimization goal is the rate-distortion performance of speech under different packet loss rate channel, where both the classification cross entropy and bandwidth consumption are considered. For fine-tuning the large speech model, the reconstruction loss is defined as:
\begin{equation}
    \begin{split}
    \mathcal{L}_{recon} &= \frac{1}{N}\sum_{n=1}^{N} \Bigg( \text{CE}(l_{n,1}, V_{n,1})+\\
    &\frac{1}{\sum_{k=2}^{N_{Q}+1}\lambda_k}\sum_{k=2}^{N_Q+1}\lambda_k\text{CE}(l_{n,k},V_{n,k}) \Bigg),
    \end{split}
\end{equation}
where $\text{CE}(\cdot)$ denotes the cross entropy loss function, and $\lambda_k$ denotes the weights of semantic tokens and acoustic tokens. 
To jointly optimize for reconstruction quality and bitrate efficiency, we incorporate a bitrate regularization term based on the number of transmitted tokens. Combining  with the bitrate constraint in (\ref{loss_bitrate}), the final loss function is as follows:
\begin{equation}
    \mathcal{L} = \mathcal{L}_{\text{recon}} + \gamma \sum_{i=1}^{N_Q} m_{n_i},
\end{equation}
where $\gamma$ is a hyperparameter that controls the trade-off between reconstruction quality and bandwidth usage.

\begin{figure*}[!ht]
    \centering
    \begin{subfigure}{0.45\textwidth}
        \centering
        \includegraphics[width=\textwidth]{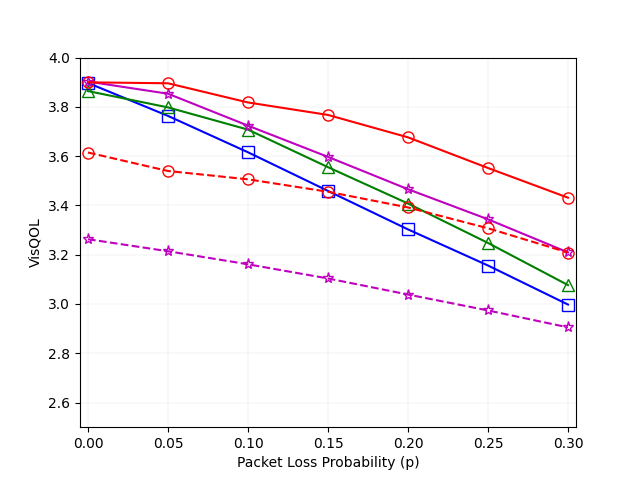}
        \caption{}
    \end{subfigure}
    \begin{subfigure}{0.45\textwidth}
        \centering
        \includegraphics[width=\textwidth]{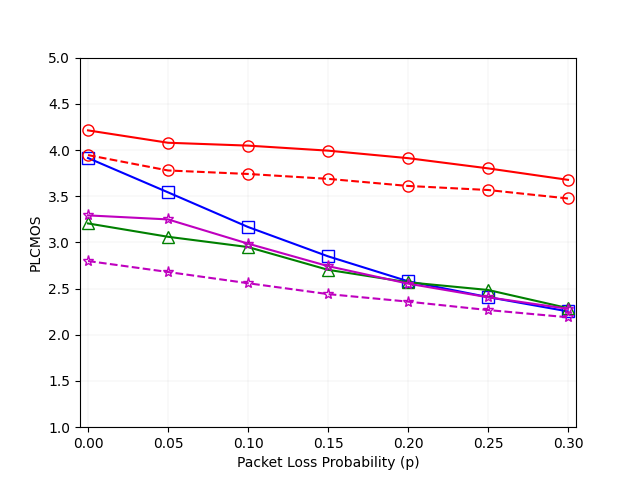}
        \caption{}
    \end{subfigure}
    \begin{subfigure}{0.45\textwidth}
        \centering
        \includegraphics[width=\textwidth]{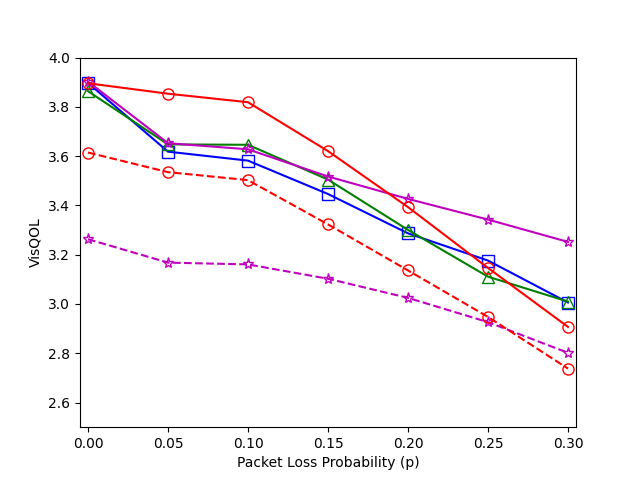}
        \caption{}
    \end{subfigure}
    \begin{subfigure}{0.45\textwidth}
        \centering
        \includegraphics[width=\textwidth]{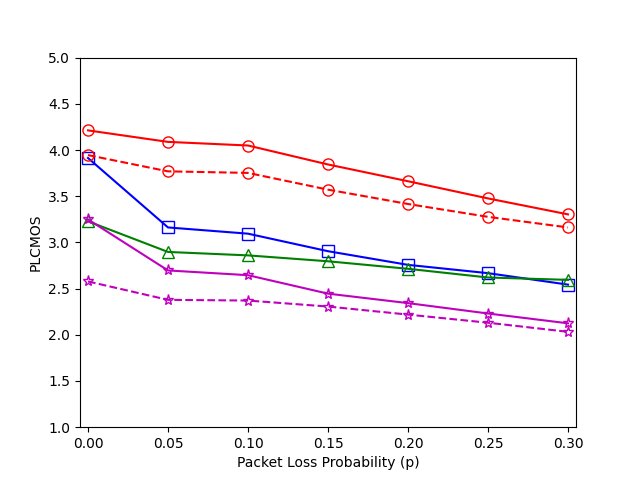}
        \caption{}
    \end{subfigure}
    
    \vspace{-0.05em}
    
    \begin{subfigure}{0.6\textwidth}
        \centering
\includegraphics[width=\textwidth]{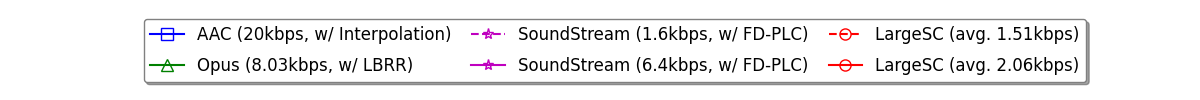}
    \end{subfigure}
    
       \caption{Audio quality performance of the proposed system and competing methods under packet loss network:  (a) VisQOL, (b) PLCMOS under random uniform channel, (c) VisQOL, (d) PLCMOS under Gilbert-Elliott channel model.}
       \label{fig:5}
\end{figure*}

\section{Simulation Results}
\label{sec:IV}
In this section, the simulation results are presented to evaluate the advantages of our proposed method over baseline methods.
\subsection{Datasets and Settings}
\subsubsection{Datasets}
To train and test the semantic encoder and finetune the large model, we use the LibriSpeech dataset \cite{dataset} as the source of transmission. This dataset is an unlabeled speech collection with a sampling rate of 16kHz. We resample the speech to 24kHz to meet the requirements of the large model. The train-clean-100 part is used to train and finetune the model, while the test-clean-100 part is used to test. To fine tune the large model, timestamp aligned text is obtained through speech recognition. To further evaluate the performance of semantic coding under untrained conditions, we additionally employ the Common Voice \cite{ardila2020common} dataset as a zero-shot test set.

\subsubsection{Training Details}
The proposed system has two training steps, using the AdamW optimizer. The first step is to finetune the Moshi large model. We set the
hyperparameter $N_Q=8$ for Mimi and $r=8$ for LoRA, with a learning rate $1\times10^{-6}$. The batch size is set as 16 with 100 steps gradient accumulation. The second step is end-to-end training of the large model with the adaptive controller under different packet loss rate. Taking inspiration from \cite{soundstream}, in order to maintain performance at low packet loss rates, we set the probability of drop out to 0.75, which means that there is a 75\% chance of packet loss for data within a batch. In addition, to support variable bitrates, following \cite{VBRVQ}, during training, we set the number of available quantizers $L$ to be randomly sampled within the range of 1 to 16.

\subsubsection{Packet Loss Channel Settings} During the training phase, in order to improve training efficiency, we set each speech token to be independent of each other in terms of packet loss, which is a random uniform packet loss model. 
In the testing phase, the random uniform loss model is considered. In addition, the Gilbert-Elliott (GE) model with two-state Markov is used to evaluate the performance of the model in burst packet loss networks. The average packet loss rate $p$ ranges from 0 to 30\%.
\subsubsection{Baselines} Advanced audio coding (AAC) \cite{aac} and Opus \cite{opus} are used as traditional speech coding methods for comparison. Soundstream is used as a neural network audio coding method in \cite{soundspring}.  When comparing rate distortion performance, the bit rate range of AAC is 4.7kbps-20.5kbps, while the bit rate range of Opus is 5.4kbps to 8.0kbps. The bitrate range of Soundstream is 1.6kbps to 6.4kbps. 
In packet loss networks, we compare baseline methods with similar performance at $p=0$ and the SoundStream method under a similar bandwidth constraint.
AAC uses frame interpolation as a packet loss compensation method. Opus uses in-band FEC method, low bit-rate redundancy (LBRR). Soundstream uses feature-domain packet loss concealment method, FD-PLC \cite{FD_PLC}.

\subsubsection{Performance Metrics}
We use objective metrics VisQOL \cite{chinen2020visqol} and subjective estimation metrics UTMOS \cite{saeki2022utmos} and PLCMOS \cite{diener2023plcmos} to evaluate the quality of reconstructed speech. VisQOL reflects the distortion of reconstructed speech and reference speech signals, while subjective indicators model users' perceptual preferences and auditory experiences through deep learning.
UTMOS is used to evaluate lossless transmission, while PLCMOS is used to evaluate the performance of packet loss concealment algorithms. In addition, Word Error Rate (WER) \cite{WER} is used to measure semantic errors of speech, where the transcripts are obtaied by SenseVoice \cite{an2024funaudiollm}. WER quantifies the proportion of substitutions, deletions, and insertions between the predicted and reference transcripts.  
To further assess prosodic naturalness, we compute the $\text{log} $ $F_0$ Root Mean Squared Error ($\text{log}F_0$ RMSE)  \cite{logF0RMSE}, which measures the deviation of the reconstructed pitch contour from the reference on a logarithmic scale, reflecting pitch stability and smoothness.

\begin{figure*}[!ht]
    \centering
    \begin{subfigure}{0.45\textwidth}
        \centering
        \includegraphics[width=\textwidth]{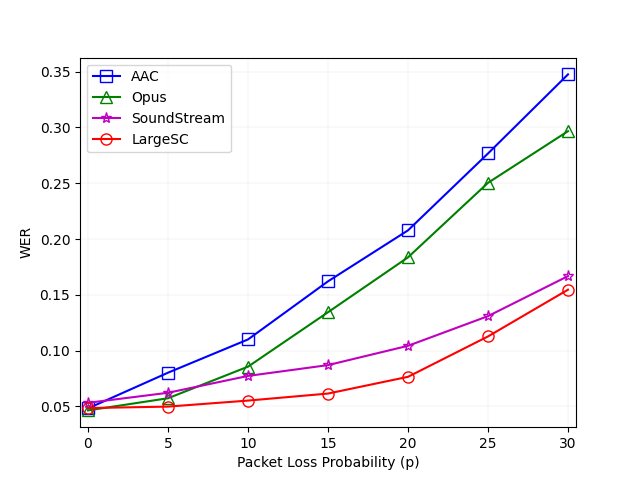}
        \caption{}
    \end{subfigure}
    \begin{subfigure}{0.45\textwidth}
        \centering
        \includegraphics[width=\textwidth]{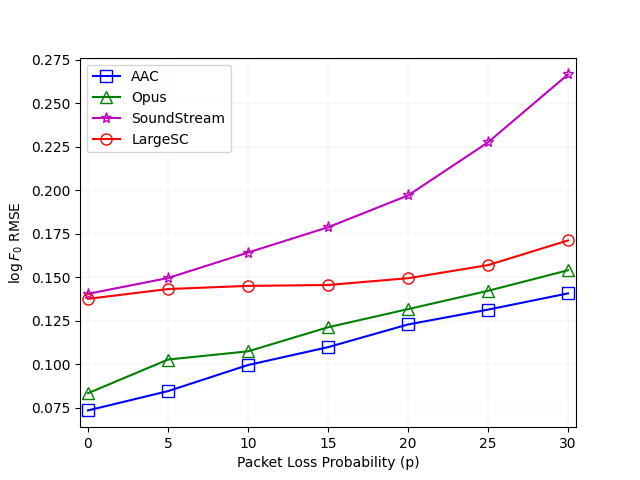}
        \caption{}
    \end{subfigure}
    \caption{Word error rate (WER) and $\text{log}F_0$ RMSE performance of the proposed system and competing methods under random uniform channel: (a) WER, (b) $\text{log}F_0~RMSE$.}
    \label{fig_wer_f0}
\end{figure*}

\subsection{Performance of Adaptive Bitrate Controller}
The performance verus bitrate of the proposed speech semantic method and baselines are shown in Fig. \ref{fig_4}. As with our method, the frame size for AAC is set to  80ms and Opus is set to 60ms. We compare the speech semantic coding method in \cite{soundspring}, and the bitrate  of Soundstream is adjusted by the used the number of codebooks. As shown in Fig. \ref{fig_4}, our method has an advantage in reconstruction quality compared to the baseline method at low bit rates. 
As shown in Fig.4 (a), when the VisQOL value is about 3.85, our method can save about 95\% bandwidth compared to the AAC method, 86\% bandwidth compared to the Opus method, and 83\% bandwidth compared to the Soundstream method. As shown in Fig. \ref{fig_4} (b), our method can achieve higher UTMOS performance within the same bit rate range. 
Therefore, our semantic coding method maintains a high compression rate and good speech quality performance while supporting variable bit rates.

\subsection{Perceptual and Semantic Quality under Different Packet Loss Rates}
For the channel with random uniform packet loss, the quality performance of our proposed system is shown in Figs. \ref{fig:5} (a) and (b). Although our method has a lower bitrate, it outperforms baselines in terms of quality performance. We compare the two average bit rates of the proposed method under packet loss rate network, namely 1.51kbps and 2.06kbps. With the occurrence of packet loss, the performance on VisQOL is superior to baseline methods. Specifically, as packet loss occurs, LargeSC (avg. 2.06 kbps) outperforms the baseline methods in terms of VisQOL under similar lossless performance. Under comparable bandwidth conditions, LargeSC (avg. 1.51 kbps) also surpasses SoundStream (1.6 kbps) in VisQOL. For the PLCMOS metric, LargeSC outperforms the baseline methods as well, indicating that the auditory quality of speech will not experience a cliff like decline due to packet loss.

As shown in Figs. \ref{fig:5} (c) and (d), the erase channel of the GE model is used to test the anti burst loss capability of our method.
As shown in Fig. \ref{fig:5} (c), when $p<0.2$, our method (2.06kbps) has a higher VisQOL value relative to baselines, indicating that our method is closer to the reference true value speech. when $p<0.3$, LargeSC (avg. 1.51kbps) outperforms the Soundstream (1.6kbps). At higher average packet loss rates, LargeSC (avg. 2.06kbps) differs from the Opus method with in-band FEC in terms of VisQOL metrics. This is because our method's packet loss prediction relies solely on previously received frames. As shown in Fig. \ref{fig:5} (d), LargeSC (avg. 2.06kbps) and LargeSC (avg. 1.51kbps) have higher PLCMOS compared to the baseline methods when packet loss occurs.
This indicates that our methods still maintain good listening experience when VisQOL is low.
Therefore, our methods maintain better resistance to packet loss at different packet loss rates.

Fig. \ref{fig_wer_f0} (a) illustrates the WER performance of different codes under increasing packet loss rates. As the loss rate grows, all methods experience a monotonic increase in WER; however, the proposed LargeSC consistently achieves the lowest WER across all evaluated packet loss probabilities. When $p=0.3$, LargeSC reaches a WER of approximately 0.15, significantly lower than SoundStream ($\approx0.17$), Opus ($\approx0.29$), and AAC ($\approx0.35$). These results indicate that the lexical intelligibility of the reconstructed speech remains more robust under the proposed model. In practical terms, LargeSC preserves more phonetic and linguistic information during packet loss, enabling downstream ASR systems to generate more accurate transcripts compared to the baseline codecs.

Fig. \ref{fig_wer_f0} (b) shows the $\text{log}F_0$ RMSE of different methods under different packet loss rates. The proposed LargeSC maintains consistently lower $\text{log}F_0$ RMSE than SoundStream across all evaluated loss rates, indicating better preservation of prosodic structure. Although AAC and Opus achieve slightly lower RMSE values under low-loss conditions, their performance degrades more rapidly as the packet loss rate grows. The $\text{log}F_0$ RMSE performance of deep learning based methods such as LargeSC and Soundstream is inferior to traditional methods. This is primarily because conventional waveform codecs typically apply deterministic filterbanks and linear prediction modules that explicitly preserve low-level spectral and pitch-related structures. Moreover, LargeSC is optimized for perceptual quality and packet loss resilience, rather than strict frame-level spectral fidelity.
\begin{figure*}[!ht]
    \centering
    \begin{subfigure}{0.45\textwidth}
        \centering
        \includegraphics[width=\textwidth]{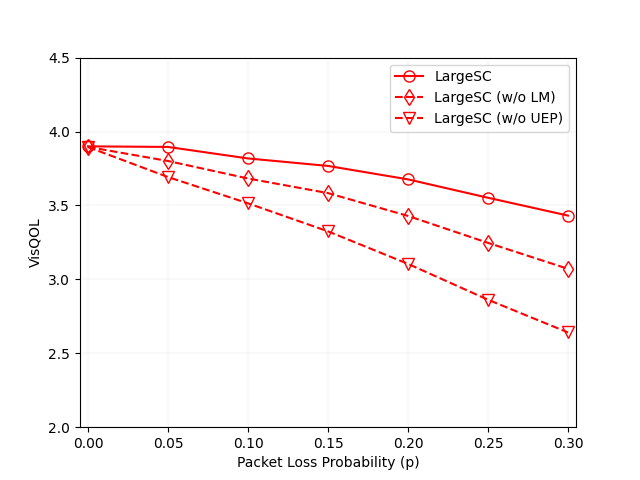}
        \caption{}
    \end{subfigure}
    \begin{subfigure}{0.45\textwidth}
        \centering
        \includegraphics[width=\textwidth]{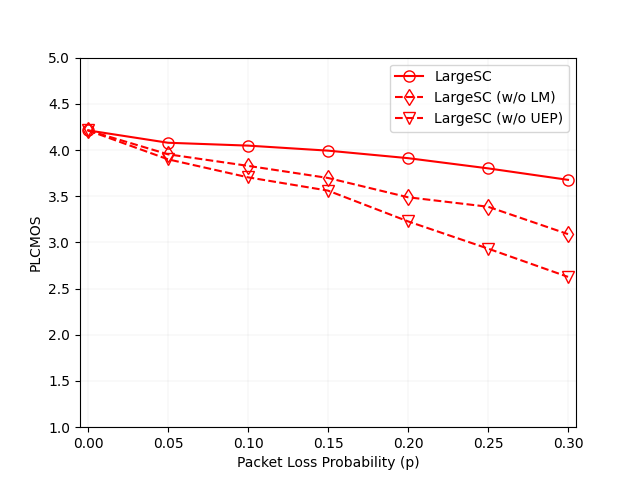}
        \caption{}
    \end{subfigure}
   
       \caption{The ablation results of unequal error protection and large model in our method under uniform random loss channel: (a) VisQOL, (b) PLCMOS of audio quality performance.
       }
       \label{fig_without}
\end{figure*}

\subsection{Ablation Analysis and Generalization Evaluation}
\subsubsection{Ablation Analysis}
Fig. \ref{fig_without} presents the ablation results of LargeSC under different packet loss rates by removing (i) the large speech model (w/o LM) and (ii) the unequal error protection capability of the adaptive rate controller module (w/o UEP). Both perceptual metrics, VisQOL and PLCMOS, degrade as packet loss increases, but the degree of degradation varies significantly across model variants. The full LargeSC model achieves the highest performance, demonstrating better robustness. Removing the LM reduces VisQOL and PLCMOS, indicating that the LM helps recover plausible acoustic information when frames are missing. The LargeSC (w/o UEP) shows greater degradation because the loss of perceptually important tokens is not protected, leading to greater distortion especially at higher loss rates. These ablation results demonstrate that both LM and UEP contribute substantially to the robustness of LargeSC, but UEP has a more pronounced impact under higher packet loss conditions, while the LM prior mainly enhances perceptual smoothness and continuity across all loss levels.

\subsubsection{Generalization Performance}
\begin{figure*}[!ht]
 \centering
     \begin{subfigure}{0.45\textwidth}
        \centering
        \includegraphics[width=\textwidth]{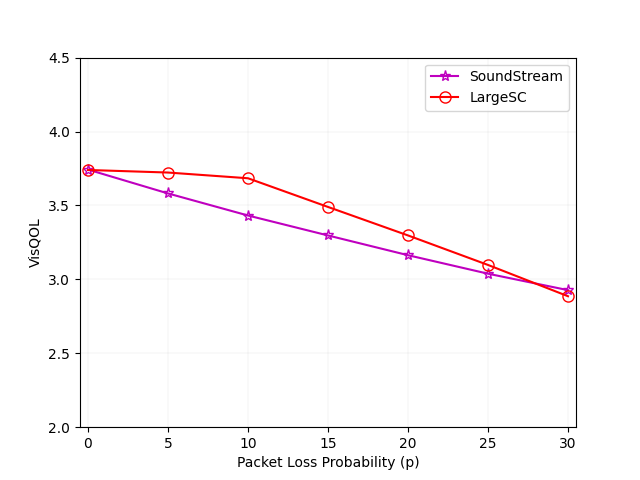}
        \caption{}
    \end{subfigure}
    \begin{subfigure}{0.45\textwidth}
        \centering
        \includegraphics[width=\textwidth]{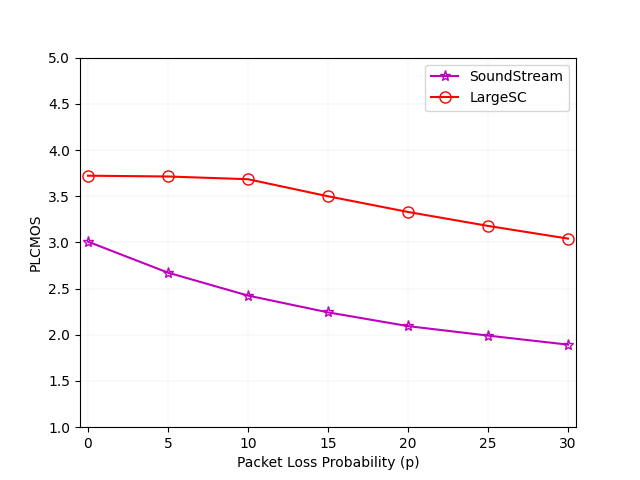}
        \caption{}
    \end{subfigure}
    \caption{Comparison of generalization reconstruction audio quality on untrained Commom Voice dataset: (a) VisQOL, (b) PLCMOS of audio quality performance.
       }
       \label{fig_commvoice}
\end{figure*}

As shown in Fig. \ref{fig_commvoice}, the evaluation on the unseen Common Voice dataset is given. For the VisQOL and PLCMOS metrics, the performance of LargeSC and Soundstream trained on LibriSpeech decreases on the Common Voice dataset. However, LargeSC consistently outperforms SoundStream across all loss rates. LargeSC maintains higher perceptual similarity and overall listening quality, indicating that its learned representations generalize more effectively to new speakers and recording conditions not seen during training. This demonstrates that the larger model exhibits stronger generalization capability, benefiting from its richer acoustic prior and greater modeling capacity. The performance gap becomes more pronounced at moderate to high packet loss levels, where SoundStream degrades rapidly. These results highlight the inherent robustness and cross-dataset adaptability brought by scaling the model capacity.

\subsection{Visualization Results in Real-Time Transmission}
\begin{figure*}[!ht]
    \centering
    \begin{subfigure}{0.45\textwidth}
        \centering
        \includegraphics[width=\textwidth]{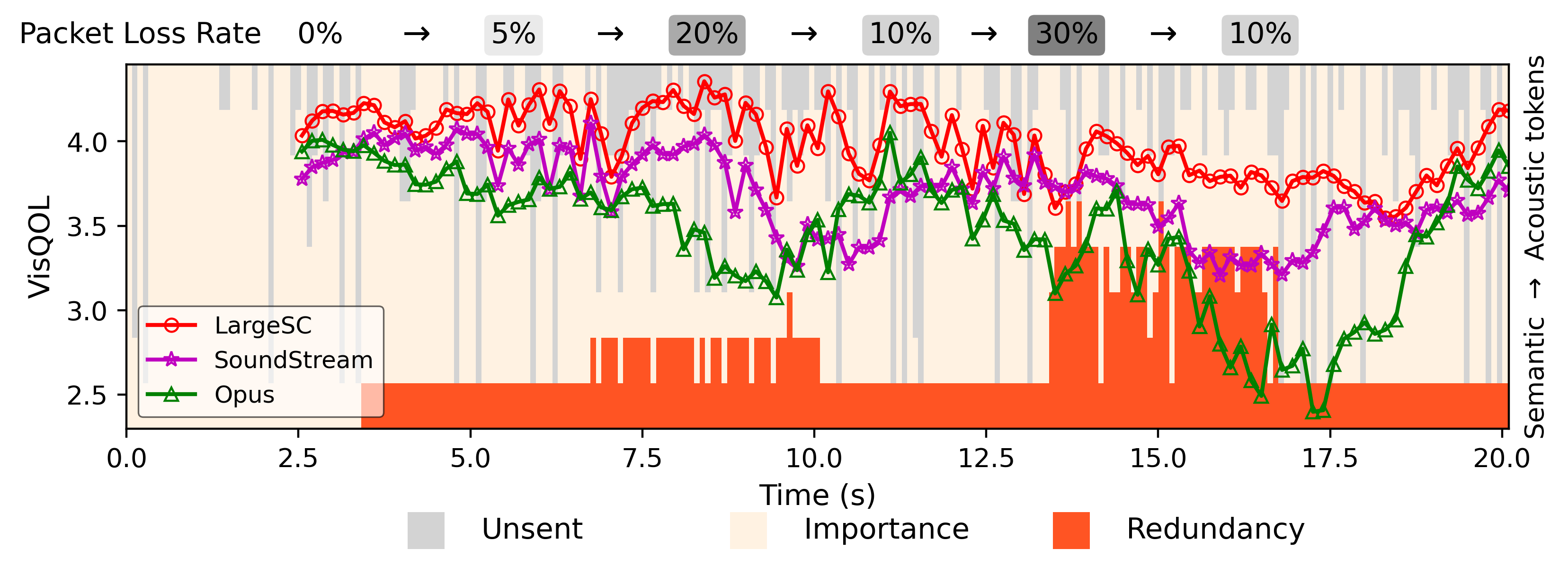}
        \caption{}
    \end{subfigure}
    \begin{subfigure}{0.45\textwidth}
        \centering
        \includegraphics[width=\textwidth]{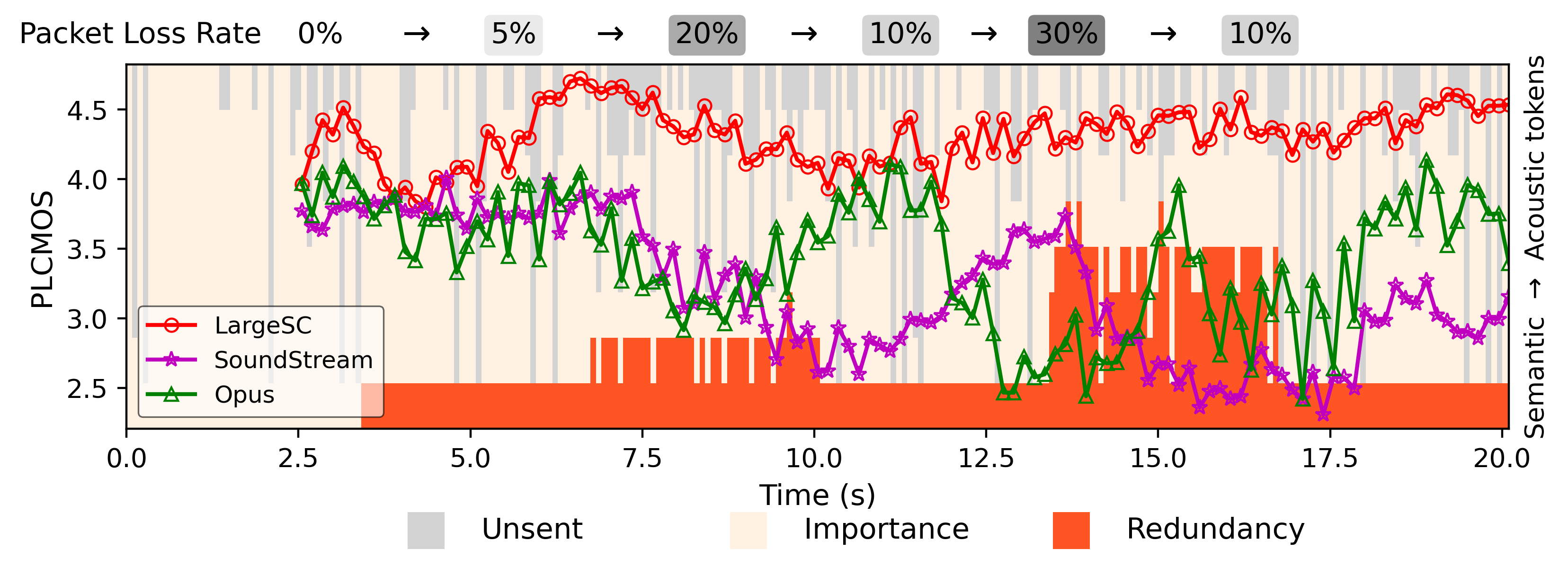}
        \caption{}
    \end{subfigure}
    \begin{subfigure}{0.45\textwidth}
        \centering
        \includegraphics[width=\textwidth]{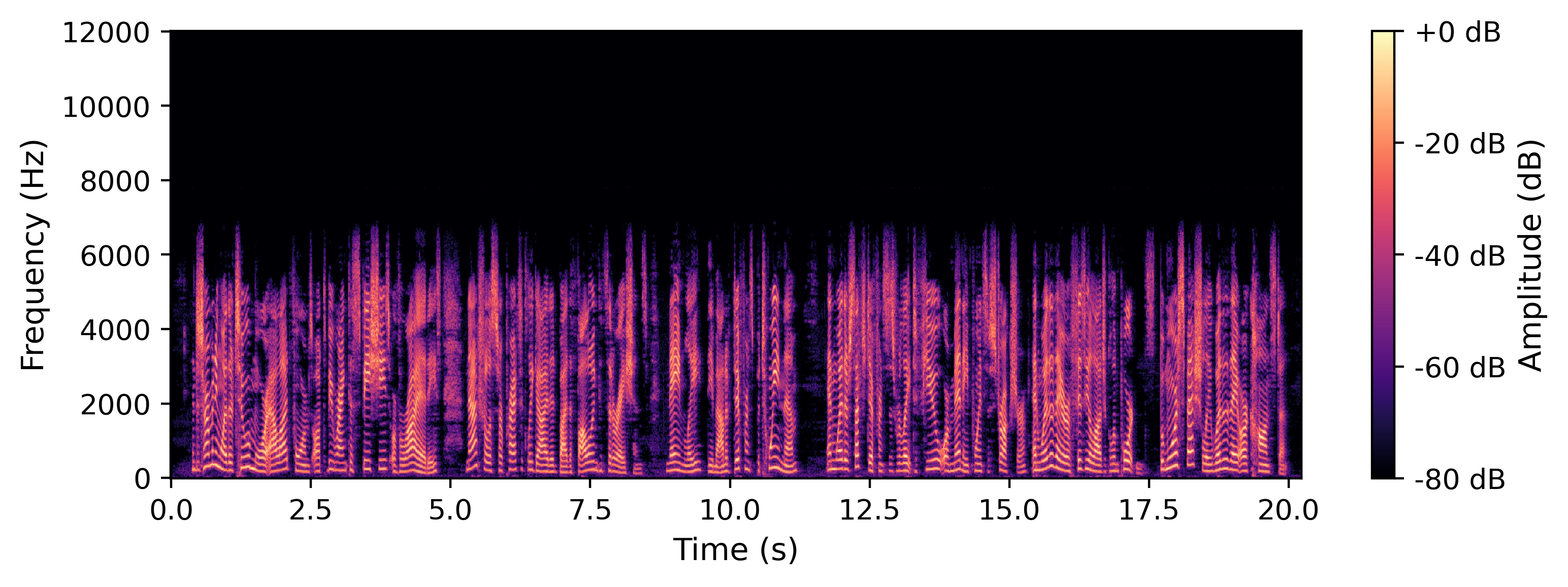}
        \caption{}
    \end{subfigure}
    \begin{subfigure}{0.45\textwidth}
        \centering
        \includegraphics[width=\textwidth]{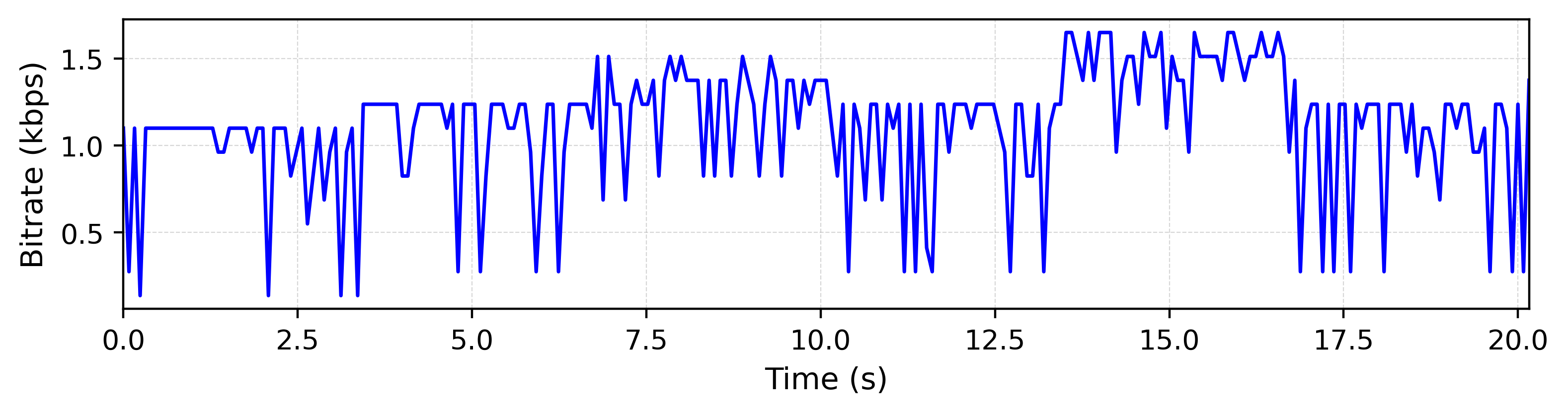}
        \caption{}
    \end{subfigure}
    \caption{Visualization of the real-time transmission process. Real-time (a) VisQOL and (b) PLCMOS scores using a 2.5-s sliding window under varying packet loss rates, with the background indicating the decisions of  adaptive controller: gray regions correspond to unsent tokens, yellow regions denote tokens predicted as important, and red regions represent redundant tokens selected for transmission. (c) Spectrogram of the original speech. (d) Adaptive bitrate variation of LargeSC during transmission. }
    \label{fig_visual}
\end{figure*}

Fig. \ref{fig_visual} visualizes the real-time transmission process of a speech sample under varying packet loss rates. Fig. \ref{fig_visual} (c) shows the spectrogram of the original speech, serving as the reference for evaluating reconstruction quality. In the transmission simulation, the packet loss rate changes dynamically across several phases (0\%, 5\%, 20\%, 10\%, 30\%, and back to 10\%) and the mask output by the adaptive controller is represented in the background of Fig. \ref{fig_visual}.

As shown in the VisQOL and PLCMOS curves (Fig. \ref{fig_visual} (a) and (b)), LargeSC maintains higher perceptual quality than SoundStream and Opus throughout the entire transmission period. The learned mask (importance and redundancy bars) adapts closely to the speech content in Fig. \ref{fig_visual} (c): frames containing important phonetic or prosodic structures are more likely to be preserved, while redundant frames are masked more aggressively. When the packet loss probability increases, the model expands the set of frames marked as important, ensuring that critical semantic information continues to be transmitted, thus stabilizes perceptual quality under adverse network conditions. 

The bitrate trajectory in Fig. \ref{fig_visual} (d) further illustrates this adaptive behavior. The bitrate fluctuates between approximately 0.4–1.6 kbps depending on the instantaneous importance of frames and the current packet loss rate. During low-loss intervals at the beginning, LargeSC reduces redundant frames, keeping the bitrate relatively low. When the channel becomes unreliable, the bitrate increases as the masking mechanism assigns a higher proportion of frames to important tokens, mitigating the impact of lost packets.

Although VisQOL and PLCMOS temporarily decline during high-loss periods, LargeSC remains consistently above the baseline codecs, offering smoother quality transitions and more robust reconstruction. The results demonstrate that the learned masking strategy effectively couples token selection with both speech semantics and channel state, achieving adaptive bitrate control while maintaining strong perceptual robustness in real-time transmission scenarios.

\subsection{End-to-End Latency Analysis}

For a real-time speech semantic communication system, the end-to-end transmission delay mainly consists of the following components: the duration of frames involved in coding, $T_{\text{Context}}$, coding computation time, $T_{\text{Coder}}$, and transmission time, $T_{\text{Transmit}}$. 
\begin{equation}
    T_{\text{total}} = T_{\text{Context}}+ T_{\text{Coder}} + T_{\text{Transmit}}.
\end{equation}
Specifically, for our proposed system, the total delay $T_{\text{total}}$ can be expressed as follow:
\begin{equation}
    T_{\text{total}} = T_{\text{Context}}+ T_{\text{Coder}} + T_{\text{RA}} +  T_{\text{LM}} + T_{\text{transmit}},
\end{equation}
where $T_{\text{RA}}$ represents the calculation time of adaptive controller, and $T_{\text{LM}}$ represents the prediction calculation time of the large model. The $T_{\text{LM}}$ is determined by the number of received tokens $\tilde{x}_n$ and the time of predict a token, $T_{\text{token}}$, which can be expressed as:
\begin{equation}
    T_{\text{LM}} = \mathbb{E}[N_{\tilde{x}_n} T_{\text{token}}],
\end{equation}
where $N_{\tilde{x}_n}$ represents the number of received tokens, determined by the compression performance of the adaptive controller and the current packet loss rate $p$.

As shown in Table. \ref{tab:configurations}, we evaluate the end-to-end latency of several speech communication systems without considering the transmission delay $T_{\text{Transmit}}$. All experiments, except for the LargeSC(low) configuration, are conducted on an Intel Xeon Gold 6342 CPU with an NVIDIA A100 GPU. The LargeSC(low) variant is measured separately on an Intel i7-12700 CPU with an NVIDIA RTX 3080 GPU to reflect its performance on a more consumer-level hardware platform.
Opus exhibits the lowest delay due to its lightweight traditional codec design and minimal contextual dependency. SoundStream also maintains low latency, benefiting from a feed-forward architecture with short receptive fields, though its coding time is longer than Opus. SyncSC, which adopts a speech-to-text and text-to-speech pipeline, incurs substantial delay because large ASR and TTS models require long context windows and introduce significant coding overhead in both directions. Although our method does not target minimal latency and employs a large autoregressive model, the overall delay remains within a reasonable range. Although the large model has high computational complexity, the causal autoregressive inference scheme enables fast step-by-step generation in streaming mode, as the model only needs to predict a small amount of tokens at each step rather than processing long future context.
Therefore, our method shows strong potential in real-time communication scenarios. We believe that further reduction in latency can be achieved through methods such as quantization, pruning, and distillation of the large model.

\begin{table}[t]
\centering
\caption{FLOPs and parameter counts of each module in the proposed system.}
\label{tab:flops_params}
\begin{tabular}{lcc}
\toprule
\textbf{Module} & \textbf{GFLOPs} & \textbf{Params (M)} \\
\midrule
Semantic Encoder            & 0.476  & 39.392 \\
Adaptive Controller            & 0.013  & 0.517  \\
Large Speech Model     & 2.358  & 7690.269 \\
Semantic Decoder            & 0.495  & 39.917 \\
\midrule
\textbf{Total}          & 3.343 & 7770.095 \\
\bottomrule
\end{tabular}

\end{table}

\subsection{Computational Complexity Analysis}
As shown in \ref{tab:flops_params}, the computational cost and parameter scale of each component of LargeSC are summarized. The overall complexity is dominated by the large model, which contains about 7B parameters and accounts for 2.36 GFLOPs per one-second speech input. This is expected, as the large model is responsible for high-level semantic prediction and thus requires a large-capacity transformer backbone. In contrast, the semantic codec and adaptive controller are significantly lighter. Therefore, further optimization toward faster inference is necessary for practical deployment, for example by adopting efficient acceleration techniques such as token-efficient modeling, structural compression, and lightweight architectures.

\subsection{Communication Overhead Analysis}
The proposed system transmits discrete speech tokens at a variable payload bitrate between 550 bps and 2.06 kbps, depending on the number of active quantizers selected by the adaptive controller. Beyond this payload, practical deployments must also account for protocol headers and control signaling. Assuming that $N_{\text{pkt}}$ packets per second are generated, the total bitrate can be approximated as
\begin{equation}
    R_{\text{total}} 
\approx R_{\text{payload}} + N_{\text{pkt}} \cdot R_{\text{header}} + R_{\text{ctrl}},
\end{equation}
where $R_{\text{payload}}$ denotes the semantic payload rate, $R_{\text{header}}$ is the number of bits per packet used by the transport and network headers (eg. RTP/UDP/IP), and $R_{\text{ctrl}}$ accounts for additional control information such as periodic feedback of the estimated packet loss probability $p$ used by the adaptive controller. In the configuration of LargeSC, speech packets are sent every 160ms, resulting in approximately $N_{pkt} = 6.25$ packets per second. Assuming a typical RTP/UDP/IP header of 40 bytes per packet, the header overhead is $N_{pkt} * R_{header} =2\text{kbps}$. The packet-loss probability 
$p$ is reported every 100 ms using a single-byte representation, the control signaling overhead is
$R_{ctrl} = 80\text{bps}$, Therefore, the total bandwidth is
$R_{total} \approx R_{payload}+2.08 \approx 2.6\text{–}4.2~\text{kbps}$. This total rate remains in the low-kbps range, significantly lower than conventional speech codecs under comparable conditions.

\begin{table}[!t]
  \caption{End-to-end latency comparison of different methods.}
  \label{tab:configurations}
  \centering
\resizebox{0.49\textwidth}{!}{
  \begin{tabular}{lccccc}
    \toprule
    Method & Sample rate (kHz) & Bitrate (bps) & $T_{\text{Coder}}$ (ms) &  $T_{\text{Context}}$ (s) & Delay (s) \\
    \midrule
    Opus & 24 & 8k & 0.00077& 0.06& 0.061\\
    Soundstream & 24 & 6.4k & 0.044 & 0.013 &  0.057 \\
    SyncSC & 16 & 55.37 & 0.18 & 1.68 & 1.86 \\
    LargeSC & 24 & 2.06k & 0.30 & 0.16 & 0.46 \\
    LargeSC(low) & 24& 2.06k& 0.32 & 0.16 & 0.48 \\
    \bottomrule
  \end{tabular}
  }
  \begin{flushleft}
    \begin{itemize}
    \item[*] The above experimental results except LargeSC(low) are conducted on the platform of Intel Xeon Gold 6342 (CPU) and one NVIDIA A100 (GPU).
    \item[*] The experimental results of LargeSC(low) are conducted on the platform of Intel i7-12700 (CPU) and one NVIDIA RTX 3080 (GPU).
    \end{itemize}
    \end{flushleft}
\end{table}

\section{Discussion}
\label{sec:V}
\subsection{Applications}
The proposed large speech model enabled semantic communication system features content-adaptive compression and unequal error protection tailored to channel conditions. Compared to conventional methods, it achieves lower transmission bandwidth, making it suitable for bandwidth-constrained scenarios such as video conferencing and satellite communications. The system also demonstrates the potential for real-time communication under high packet loss rates, addressing connectivity needs in remote or underserved areas.

The token-based transmission scheme is compatible with existing digital communication protocols and operates at a low bitrate. As the codebook is shared only between the transmitter and receiver,the approach is well-suited for secure communication. By leveraging the generative capabilities and rich semantic representations of large language models, the system exhibits good generalization to diverse user speech inputs without relying on task-specific datasets, indicating potential for deployment in large-scale multi-user semantic communication systems.
Therefore, the proposed method provides improved communication efficiency compared to traditional approaches and holds strong potential for applications in low-bandwidth scenarios.

\subsection{Limitations and Challenges}
\subsubsection{Computational Requirements} 
The proposed method leverages a large-scale generative model with approximately 7 billion parameters, enabling enhanced semantic representation and generation capabilities. While this design offers substantial advantages over conventional approaches, it also introduces significant computational demands during inference. This raises important deployment considerations, particularly for real-time communication scenarios. Cloud-based or high-performance server infrastructures can provide sufficient computational resources for running such large models. However, their use may introduce additional network transmission and queuing delays, which could counteract the low-latency benefits achieved by the proposed system. At the same time, our simulation results show that LargeSC achieves an end-to-end latency of approximately 462 ms on a consumer-grade GPU (e.g., NVIDIA RTX 3080), suggesting that local deployment remains technically feasible under practical hardware conditions. Moreover, ongoing advancements in model compression—such as quantization, distillation, pruning, and inference optimization—are expected to further reduce computational requirements and make real-time on-device inference increasingly accessible.

\subsubsection{End-to-End Latency} The measured end-to-end latency is approximately 460 ms, primarily due to large model inference time and interleaving delays. The proposed system achieves lower latency than existing methods such as SyncSC, benefiting from causality and autoregressive generation. Further reductions may be achieved through architectural optimizations, including asynchronous processing or hybrid strategies that balance performance and computational load. These improvements are particularly relevant for real-time or delay-sensitive applications.

\section{Conclusion}
\label{sec:VI}
In this paper, we propose a large speech model enabled semantic communication system, LargeSC, for adaptive transmission over channels with varying packet loss rates. By employing the Mimi neural codec as speech coder, the system transforms speech into discrete tokens, thus enabling compatibility with digital semantic communication frameworks while maintaining high speech quality. An adaptive controller module is proposed to perform semantic-aware compression and channel-aware in-band unequal error protection based on current transmission conditions, enhancing both bandwidth efficiency and robustness. Furthermore, a fine-tuned Moshi speech model, adapted via LoRA, is leveraged to autoregressively predict lost or masked tokens, thereby improving reconstruction quality.
LargeSC adopts a causal, autoregressive structure, making it suitable for real-time streaming scenarios. It achieves an end-to-end latency of approximately 460 ms, demonstrating the potential for practical deployment. Therefore, the proposed system improves the efficiency and quality of speech semantic communication under bandwidth constrained and lossy channel conditions.

\ifCLASSOPTIONcaptionsoff
  \newpage
\fi

\bibliographystyle{IEEEtran}
\bibliography{IEEEabrv, bibtex/bib/reference}

\begin{IEEEbiography}[{\includegraphics[width=1in,height=1.25in,clip,keepaspectratio]{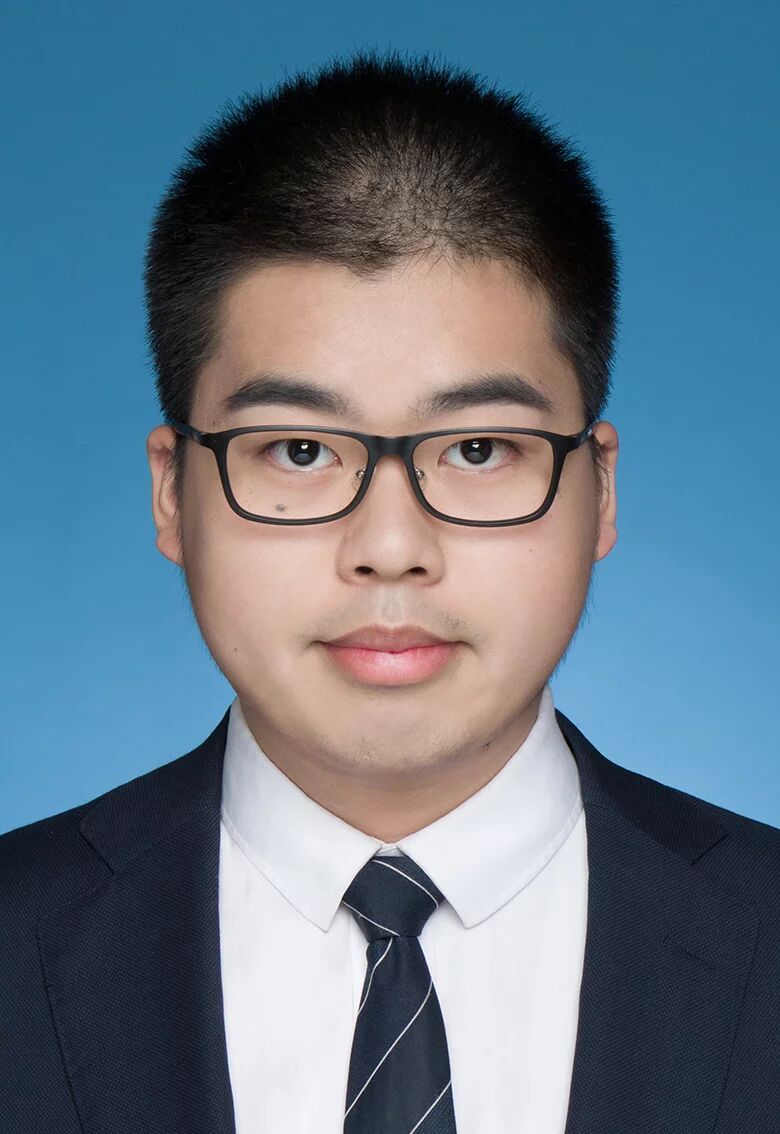}}]{Yun Tian}
received the bachelor's degree in electronic information engineering from Wuhan University, Wuhan, China, in 2021. He is currently pursuing the Ph.D. degree with the School of Electronics, Peking University, Beijing, China. His research interests include semantic communication and deep learning.

\end{IEEEbiography}

\begin{IEEEbiography}[{\includegraphics[width=1in,height=1.25in,clip,keepaspectratio]{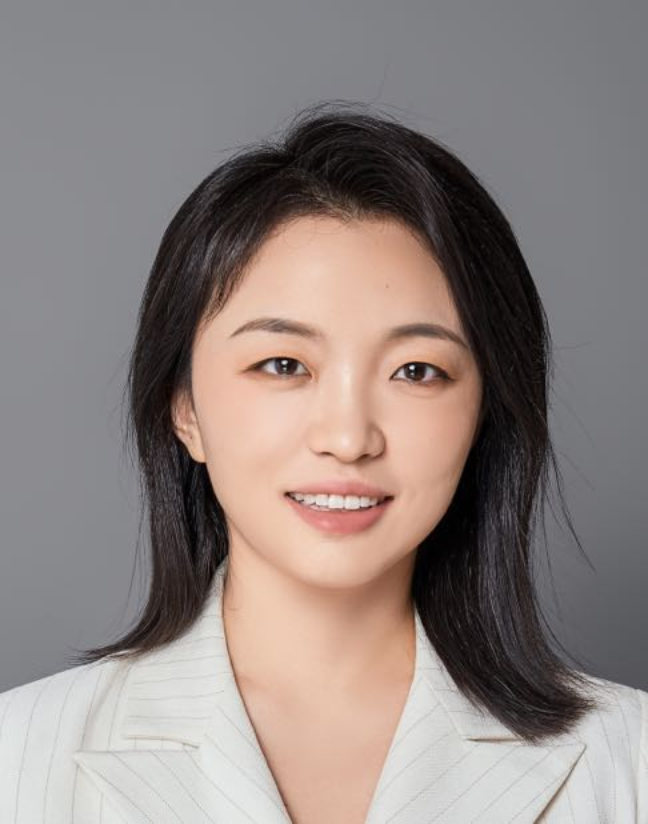}}]{Zhijin Qin}(Senior Member, IEEE)  is an Associate Professor with the Department of Electronic Engineering, Tsinghua University, Beijing, China. She was with Imperial College London, Lancaster University and Queen Mary University of London, UK, from 2016 to 2022. Her research interests include semantic communications and satellite communications. She was the recipient of the 2017 IEEE GLOBECOM Best Paper Award, 2018 IEEE Signal Processing Society (SPS) Young Author Best Paper Award, 2021 IEEE Communications Society Signal Processing for Communications Committee Early Achievement Award, 2022 IEEE Communications Society Fred W. Ellersick Prize, 2023 IEEE ICC Best Paper Award and 2023 IEEE SPS Best Paper Award. She served as an Area Editor of IEEE JSAC Series, an Associate Editor of IEEE Transactions on Communications, IEEE Transactions on Cognitive Communications and Networking. She is serving as an Area Editor of IEEE Communications Letters, an Associate Editor of IEEE Signal Processing Magazine and IEEE Transactions on Wireless Communications.
\end{IEEEbiography}

\begin{IEEEbiography}[{\includegraphics[width=1in,height=1.25in,clip,keepaspectratio]{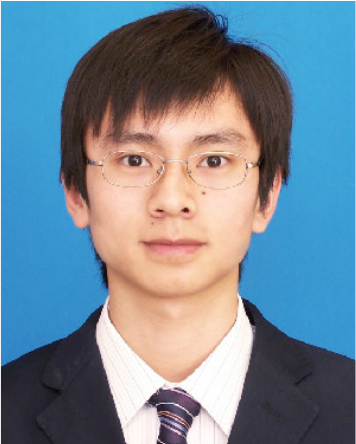}}]{Guocheng Lv} received the B.S. and M.S. degrees
from Peking University, Beijing, China, in 2006 and
2009, respectively. He is currently a Senior Engineer
with the School of Electronics, Peking University.
His research interests include satellite communication, physical-layer modem, and non-orthogonal
multiple access.
\end{IEEEbiography}

\begin{IEEEbiography}[{\includegraphics[width=1in,height=1.25in,clip,keepaspectratio]{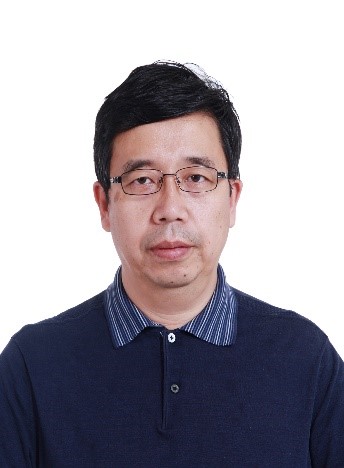}}]{Ye Jin} received the B.E. and M.S. degrees from Peking University, Beijing, China, in 1986 and
1989, respectively. He is currently a Professor with
the Institute of Modern Communications, Peking
University. He has been the Principal Investigator of over
30 funded research projects. His general research
interests are in the areas of satellite and wireless
communications and networking. Prof. Jin was a
recipient of the First Prize of the National Science
and Technology Progress Awards of China.
\end{IEEEbiography}

\begin{IEEEbiography}[{\includegraphics[width=1in,height=1.25in,clip,keepaspectratio]{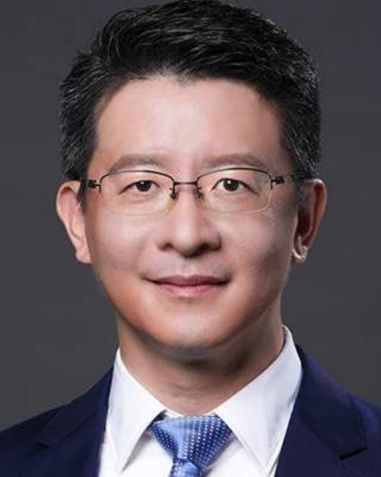}}]{Kaibin Huang} (Fellow, IEEE) received the B.Eng. and M.Eng. degrees from the National University of Singapore, and the Ph.D. degree from The University of Texas at Austin, all in electrical engineering. He is a Philip K, H, Wong Wilson K, L, Wong Professor in electrical engineering and the Department Head with the Department of Electrical and Electronic Engineering, The University of Hong Kong (HKU), Hong Kong. His work was recognized with seven Best Paper awards from the IEEE Communication Society. He is a member of the Engineering Panel of Hong Kong Research Grants Council (RGC) and a RGC Research Fellow (2021 Class). He has served on the editorial boards of five major journals in the area of wireless communications and co-edited ten journal special issues. He has been active in organizing international conferences such as the 2014, 2017, and 2023 editions of IEEE Globecom, a flagship conference in communication. He has been named as a Highly Cited Researcher by Clarivate in the last six years (2019-2024) and an AI 2000 Most Influential Scholar (Top 30 in Internet of Things) in 2023-2024. He was an IEEE Distinguished Lecturer, He is a Fellow of the U.S. National Academy of Inventors.
\end{IEEEbiography}

\begin{IEEEbiography}[{\includegraphics[width=1in,height=1.25in,clip,keepaspectratio]{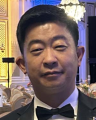}}]{Zhu Han} (Fellow, IEEE) received the B.S. degree in electronic engineering from Tsinghua University, Beijing, China, in 1997, and the M.S. and Ph.D. degrees in electrical and computer engineering from the University of Maryland, College Park, MD, USA, in 1999 and 2003, respectively. From 2000 to 2002, he was an R\&D Engineer of JDSU, Germantown, Maryland. From 2003 to 2006, he was a Research Associate with the University of Maryland. From 2006 to 2008, he was an Assistant Professor with Boise State University, Idaho. He is currently a John and Rebecca Moores Professor with the Electrical and Computer Engineering Department as well as in the Computer Science Department with the University of Houston, Texas. His main research targets on the novel game-theory related concepts critical to enabling efficient and distributive use of wireless networks with limited resources. His other research interests include wireless resource allocation and management, wireless communications and networking, quantum computing, data science, smart grid, carbon neutralization, security and privacy. He was the recipient of an NSF Career Award in 2010, the Fred W. Ellersick Prize of the IEEE Communication Society in 2011, the EURASIP Best Paper Award for the Journal on Advances in Signal Processing in 2015, IEEE Leonard G. Abraham Prize in the field of Communications Systems (best paper award in IEEE JSAC) in 2016, IEEE Vehicular Technology Society 2022 Best Land Transportation Paper Award, and several best paper awards in IEEE conferences. He was an IEEE Communications Society Distinguished Lecturer from 2015 to 2018 and ACM Distinguished Speaker from 2022 to 2025, AAAS Fellow since 2019, and ACM Fellow since 2024. He is a 1\% highly cited Researcher since 2017 according to Web of Science. He is also the winner of the 2021 IEEE Kiyo Tomiyasu Award (an IEEE Field Award), for outstanding early to mid-career contributions to technologies holding the promise of innovative applications, with the following citation: “for contributions to game theory and distributed management of autonomous communication networks.”
\end{IEEEbiography}
\end{document}